\documentclass[prl,aps,twocolumn,superscriptaddress]{revtex4}
\usepackage{amsmath,amssymb,graphicx,verbatim,hyperref}

\begin{document}

\title{Dynamically corrected gates for an exchange-only qubit}
\author{G.~T.~Hickman}
\affiliation{Department of Physics, University of Maryland Baltimore County, Baltimore, MD 21250, USA}
\author{Xin Wang}
\affiliation{Condensed Matter Theory Center, Department of Physics, University of Maryland, College Park, MD 20742, USA}
\author{J.~P.~Kestner}
\affiliation{Department of Physics, University of Maryland Baltimore County, Baltimore, MD 21250, USA}
\affiliation{Condensed Matter Theory Center, Department of Physics, University of Maryland, College Park, MD 20742, USA}
\author{S.~Das Sarma}
\affiliation{Condensed Matter Theory Center, Department of Physics, University of Maryland, College Park, MD 20742, USA}

\begin{abstract}
We provide analytical composite pulse sequences that perform dynamical decoupling concurrently with arbitrary rotations for a qubit coded in the spin state of a triple quantum dot.  The sequences are designed to respect realistic experimental constraints such as strictly nonnegative couplings.  Logical errors and leakage errors are simultaneously corrected.  A short pulse sequence is presented to compensate nuclear noise and a longer sequence is presented to simultaneously compensate nuclear and charge noise.  The capability developed in this work provides a clear prescription for combatting the relevant sources of noise that currently hinder exchange-only qubit experiments.
\end{abstract}

\maketitle

Semiconductor quantum dot spin systems offer a scalable platform for quantum computing, and have received an increasing amount of attention as the push towards a practical, large-scale array of qubits continues.  Although it is natural to consider a single localized electron spin as a qubit, with single qubit rotations controlled by applied local radio-frequency ac magnetic fields, such ESR-type single electron spin rotations have turned out to be difficult \cite{Koppens06} to implement experimentally in quantum dots. It was subsequently noted that by encoding the qubit in the spins of three electrons on a triple quantum dot, control of the exchange couplings between dots is sufficient to perform any qubit rotation \cite{DiVincenzo00}.  This ``exchange-only" qubit is highly desirable for its fast, all-electrostatic operations that avoid the complications of the local ac magnetic field control needed for single spin rotations \cite{Koppens06} or the inhomogeneous magnetic field control for singlet-triplet spin qubits \cite{Petta05}.  Recent experiments demonstrate the viability of this all-exchange approach, but also the deleterious effects of charge noise and, predominantly, hyperfine-mediated quasistatic nuclear spin fluctuations \cite{Laird10, Gaudreau12, Medford13}.

This degradation of a quantum state through the hyperfine interaction is a serious problem for quantum computation using double or triple quantum dot qubits \cite{Petta05, Maune12}. For the singlet-triplet qubit, schemes have been developed to preserve the qubit state \cite{Bluhm11,Barthel10,Medford12} and to perform gate operations while canceling errors \cite{Wang12, Kestner13}. These schemes do not apply to the exchange-only qubit though, due to an additional error channel of hyperfine-induced leakage out of the logical subspace. Error correction in this system requires a completely new approach, and to date error-limiting sequences have been proposed only for the {\sc noop}, or ``no operation," gate \cite{West12}. Such {\sc noop} operation is capable of extending the idle quantum memory as has already been demonstrated extensively for single spins \cite{Witzel07, Uhrig07, Lee08, Yang08}, but does not apply during gate operations. The main point of our work is to enable a direct method of carrying out arbitrary single qubit operations in all-exchange qubits which are dynamically decoupled from environmental noise effects.  In this Rapid Communication, we introduce a scheme for performing these rotations while canceling errors to leading order in both nuclear Overhauser field and charge fluctuations. The resulting pulse sequences are derived analytically, respect the unique physical constraints of the experimental system, and are experimentally implementable under realistic conditions.

We consider three electrons in a linear triple quantum dot system where neighboring dots are coupled by the Heisenberg exchange interaction. The qubit is encoded in the $S=1/2$ and $S^z=+1/2$ subspace as $|0\rangle=({|\!\uparrow\downarrow\uparrow\rangle}-{|\!\downarrow\uparrow\uparrow\rangle})/\sqrt{2}$ and $|1\rangle=({|\!\uparrow\downarrow\uparrow\rangle}+{|\!\downarrow\uparrow\uparrow\rangle})/\sqrt{6}-\sqrt{6}{|\!\uparrow\uparrow\downarrow\rangle}/3$ \cite{DiVincenzo00}. Quasistatic fluctuations in the nuclear Overhauser field cause the qubit states to leak to an $S=3/2$, $S^z=+1/2$ state $|Q\rangle=({|\!\uparrow\downarrow\uparrow\rangle}+{|\!\downarrow\uparrow\uparrow\rangle}+{|\!\uparrow\uparrow\downarrow\rangle})/\sqrt{3}$ via the coupling $H_{\rm hf} = \sum_j B_jS^z_j$. Here, $S^z_j$ is the spin operator in the $z$-direction for the electron in the $j$th dot, and $B_j$ is the hyperfine field. Leakage into other states, for example the ${|\!\uparrow\uparrow\uparrow\rangle}$ state, can be substantially suppressed by applying a large homogeneous Zeeman field and is therefore neglected. We also assume that Landau-Zener dynamics are suppressed as in Ref.~\cite{Medford13}, and neglect them in the remainder of this work.

In the basis $\{|0\rangle, |1\rangle, |Q\rangle\}$ we may write our Hamiltonian $H=H_{\rm c}+H_{\rm hf}$ in terms of the Gell-Mann matrices \cite{note}, $\lambda_j$, as \cite{Ladd12}
\begin{align}
H_{\rm c}=J_{12}(t)E_{12}+J_{23}(t)E_{23},
\end{align}
with
\begin{align}
E_{12}=-\frac{\lambda_3}{2}-\frac{\lambda_8}{2\sqrt{3}}, \quad E_{23}=-\frac{\sqrt{3}}{4}\lambda_1+\frac{\lambda_3}{4}-\frac{\lambda_8}{2\sqrt{3}},
\end{align}
and
\begin{align}
H_{\rm hf}=\left(\frac{\lambda_1}{2 \sqrt{3}}+\frac{\lambda_4}{\sqrt{6}}\right)\Delta_{12}+\left(\frac{\lambda_3}{3}+\frac{\sqrt{2}}{3}\lambda_6\right)\Delta_{\overline{12}},\label{eq:Hhfdef}
\end{align}
where we have defined $\Delta_{12}=B_1-B_2$ and $\Delta_{\overline{12}}=B_3-(B_1+B_2)/2$ since only the inhomogeneous part of $H_{\rm hf}$ is important \cite{Ladd12}. In the control Hamiltonian $H_{\rm c}$, $J_{ij}(t)$ is the exchange interaction between electrons in dots $i$ and $j$ which can be rapidly controlled electrostatically via the interdot detuning, $\epsilon_{ij}\left(t\right)$. $\lambda_1$, $\lambda_3$, and $\lambda_8$ act on the logical subspace as Pauli matrices $\sigma_x$, $\sigma_z$, and the identity, respectively. Therefore, $E_{12}$ and $E_{23}$ implement rotations on the Bloch sphere about axes $120^\circ$ apart, $\hat{z}$ and $\frac{\sqrt{3}}{2}\hat{x}-\frac{1}{2}\hat{z}$, respectively. Any single-qubit operation can then be composed from interleaving rotations about these two axes (or by pulsing both axes at once \cite{Shim13}).
We denote ideal rotations about these axes by $R_{12}(\phi)=\exp(-iE_{12}\phi)$ and $R_{23}(\phi)=\exp(-iE_{23}\phi)$.

There are two main sources of noise in this system. One is the nuclear noise, $H_{\rm hf}$, which causes both dephasing within the logical subspace and leakage to state $|Q\rangle$, as seen in Eq.~\eqref{eq:Hhfdef}. This is characterized by $\Delta_{12}$ and $\Delta_{\overline{12}}$. The other is the exchange noise, that is, imperfection in the control resulting for example from detuning fluctuations induced by charge noise \cite{Dial12}, $\delta\epsilon_{ij}$. Different from the nuclear noise, exchange noise is typically dependent on the strength of the control field. Thus, both of the exchange terms have the form
\begin{equation}
J\left(t\right) = J\left[\epsilon\left(t\right)\right]+\delta\epsilon \left. \frac{\partial J\left(\epsilon\right)}{\partial \epsilon}\right|_{\epsilon = \epsilon\left(t\right)},\label{eq:J12def}
\end{equation}
where $J\left[\epsilon\left(t\right)\right]$ is the desired control field and $\delta\epsilon$ arises from, \textit{e.g.}, fluctuations in the background impurity potential. In our work, we assume that $\delta\epsilon$ varies slowly enough on the timescale of one gate operation that its value can be considered constant during that period. This is justified by the fact that experimental coherent echo times are $T_{2,echo} > 0.1$ ms in GaAs systems \cite{Bluhm11} and milliseconds or even seconds in silicon-based systems \cite{Witzel10,Tyryshkin11}, compared to gate times well under a nanosecond.  In practice one sometimes produces one detuning axis for both $J_{12}$ and $J_{23}$ \cite{Medford13}, but we will not assume this, instead treating the most general case where $J_{12}$ and $J_{23}$ are controlled separately, meaning that $\delta\epsilon_{12}$ and $\delta\epsilon_{23}$ are independent noise channels. Moreover, the functional dependencies of the exchanges on detunings are to be measured experimentally. Phenomenological models exist in the literature such as $J=J_0\exp(\epsilon/\epsilon_0)$, where $J_0$ and $\epsilon_0$ are determined empirically, implying $\frac{\partial J}{\partial \epsilon}\propto J$ \cite{Khodjasteh12}. Other forms also exist \cite{Gaudreau12}. Our result, though, does not rely on any particular choice of the functional form.

With these considerations we define a na\"ive rotation of angle $\phi$ around the $z$ axis via holding a constant value of the exchange, $J$, for a time $\phi/J$,
\begin{equation}
U_{12}(J,\phi)=\exp\left\{-i\left\{\left[J+g(J)\delta\epsilon_{12}\right]E_{12}+H_{\rm hf}\right\}\frac{\phi}{J}\right\},\label{eq:U12def}
\end{equation}
and define $U_{23}(J,\phi)$ similarly. Here $g(J)$ is a shorthand notation for $\frac{\partial J\left(\epsilon\right)}{\partial \epsilon}$ evaluated at the detuning that produces exchange $J$. When there are no noise terms, the above rotation implements $R_{12}(\phi)$ exactly.  In the presence of noise, there are errors at first order related to both hyperfine interaction ($\Delta_{12}$, $\Delta_{\overline{12}}$) and exchange noise ($\delta\epsilon_{12}$, $\delta\epsilon_{34}$) \cite{note}. Our goal is to find a pulse sequence that accomplishes the desired rotation, $R_{12}(\phi)$ while canceling all leading order errors. At the same time, we also want to respect the experimental constraints that $0\leq J\leq J_\mathrm{max}$ and $\phi\ge0$ (\textit{i.e.}, time durations are nonnegative). In general, logical and leakage errors are expressed in terms of matrices $\lambda_1$ through $\lambda_7$ with coefficients that are functions of $\Delta_{12}$, $\Delta_{\overline{12}}$, $\delta\epsilon_{12}$, and $\delta\epsilon_{23}$, so one would need to solve 28 coupled highly nonlinear equations to set each of the leading order error terms to zero -- a forbidding task. However, the procedure can be substantially simplified by using symmetry considerations, specific to this qubit implementation, to eliminate most of the error terms from the outset and to allow an analytical solution that we present in this work.

In the following, we first neglect imperfections in the control of the exchange couplings and focus on canceling the hyperfine-induced noise, which recently has been argued to play a dominant role in experiments. \cite{Mehl12}. Then, using the resulting pulse sequences as building blocks, we build nested composite pulses that also cancel the exchange noise at the same time as the hyperfine noise. Since the evidence points to a noise spectral density dominated by its quasistatic component \cite{Medford13}, we design our scheme to compensate this component specifically.

While an elegant approach for leakage error correction has previously been presented in Ref.~\onlinecite{Byrd05}, we start with the alternative approach of West and Fong \cite{West12}. The idea is that hyperfine errors in an identity operation can be turned into a harmless global phase by permuting electrons between sites such that each electron sees the same average magnetic field.  This idea was used to construct spin echo sequences with delta function pulses, but we will use it to construct arbitrary corrected rotations with real pulses of finite duration.

A complete cycle of permutations is performed by $I=\left[U_{12}(J,\pi)U_{23}(J,\pi)\right]^3$. To understand how the error accumulates, we first move to a rotating frame and consider the error induced by individual pieces, then combine them by moving back to the lab frame \cite{Khodjasteh09}. The real rotations can be related to ideal ones by $U_{12}(J,\pi)=R_{12}(\pi)(1-i\Phi_{12})$ and $U_{23}(J,\pi)=R_{23}(\pi)(1-i\Phi_{23})$, with $\Phi_{12}=\sum_i a_i\lambda_i$ and $\Phi_{23}=\sum_i b_i\lambda_i$. We begin by considering a composite pulse implementing an identity up to hyperfine errors,
\begin{multline}\label{eq:6pieceU}
U_{12}(J,\pi)U_{23}(J,\pi)U_{12}(J,\pi)U_{23}(J,\pi)U_{12}(J,\pi)U_{23}(J,\pi)
\\
=I-i\Phi_{\rm tot},
\end{multline}
where the error is
\begin{multline}
\Phi_{\rm tot} = \Phi_{23}+P_1^\dagger\Phi_{12}P_1+P_2^\dagger\Phi_{23}P_2
\\
+P_3^\dagger\Phi_{12}P_3+P_4^\dagger\Phi_{23}P_4+P_5^\dagger\Phi_{12}P_5.\label{eq:Phierror}
\end{multline}
Here $P_n=\prod_{i=1}^nU_i$, and $U_i=U_{23}(J,\pi)$ for $i$ odd and $U_{12}(J,\pi)$ for $i$ even. Simple algebra gives
\begin{equation}
\Phi_{\rm tot}=-3(a_2-b_2)\lambda_2+c_8\lambda_8.\label{eq:phitot2}
\end{equation}
Equation \ref{eq:phitot2} is completely general and applies regardless of pulse shapes chosen for $U_{12}(J,\pi)$ and $U_{23}(J,\pi)$, so long as they are consistent and implement $\pi$ rotations to zeroth order. Thus we need only replace the $U_{12/23}(J,\pi)$ with composite pulses such that $a_2 = b_2$ and the identity will be free of hyperfine errors, other error terms being canceled by the cyclic permutations of Eq.~\eqref{eq:6pieceU}. (The $\lambda_8$ term is harmless, since it does not affect the qubit subspace.)

There are various ways to achieve $a_2=b_2$, but for our subsequent discussion it is most convenient to replace $U_{12/23}(J,\pi)$ above with a three-piece pulse sequence $U'_{12/23}(J,\pi)$, where
\begin{equation}
U'_{12/23}(J,\phi)=U_{12/23}(J,\phi)U_{12/23}\left(\frac{J}{2},2\pi-\phi\right)U_{12/23}(J,\phi).\label{eq:3piecePhi}
\end{equation}
This sequence has the desirable property that its error has no $\lambda_2$ component to first order in the hyperfine fields. Therefore when one implements the rotations in Eq.~\eqref{eq:6pieceU} by Eq.~\eqref{eq:3piecePhi}, $a_2=b_2=0$ and the hyperfine-induced error vanishes. The value of $J$ is unimportant as long as its value is consistent across all implementations of $U_{12}$, and similarly for $U_{23}$. (Although we have furthermore taken $J_{12}=J_{23}=J$ for simplicity, it is not necessary that the same exchange coupling be used between both pairs of dots.) We remark here that this gives an alternative dynamical decoupling scheme to that used in Ref.~\onlinecite{Medford13}. In fact, our sequence actually cancels the lowest two orders of error due to nuclear noise for this case.

Crucially, $U_{12/23}'(J,\phi)$ has another desirable property that its error is independent of $\phi$ and is thus always identical to that associated with $U_{12/23}'(J,\pi)$. Both properties are by design, and spring from intuition that $U_{12}'(J,\phi)$ should always carry the qubit through at least a full circle of rotation around the Bloch sphere, spending an equal amount of time at each pair of diametrically opposite points. Since $J_{12}$ (or $J_{23}$) performs these rotations about an axis orthogonal to the y-axis, the first-order $\lambda_2$ error component generated at each point of the evolution (not just during waiting times) is cancelled by that from the other side of the Bloch sphere and the remaining first-order errors are independent of $\phi$. (The expanded expression for Eq.~\eqref{eq:3piecePhi} given in \cite{note} confirms this intuition.)

The latter property allows us to implement a corrected nontrivial qubit rotation instead of an identity. We consider a variation of the sequence of Eq.~\eqref{eq:6pieceU}, using the composite pulses of Eq.~\eqref{eq:3piecePhi} and replacing the last $\pi$ rotation by one with a general angle,
\begin{multline}
\widetilde{U}_{12}(\phi)\equiv U_{12}'(J,\pi+\phi)U_{23}'(J,\pi)\left[U_{12}'(J,\pi)U_{23}'(J,\pi)\right]^2
\\
=R_{12}(\phi) + \mathcal{O}\left[\left(\Delta_{12}+\Delta_{\overline{12}}\right)^2\right].\label{eq:hfcorrdU12}
\end{multline}
The total sequence error remains zero to first order since the error of the last piece does not depend on $\phi$. Changing the rotation implemented by the last piece of the sequence only manifests itself in the zeroth order term since $P_6$ does not appear in the expression for total error \eqref{eq:Phierror}. To implement an $R_{23}(\phi)$ rotation one simply interchanges 12 and 23 indices. Thus, neglecting exchange errors at this stage, we have found a prescription to perform arbitrary single-qubit corrected operations. The explicit analytical form showing hyperfine-noise cancellation, the remaining $\lambda_8$ term, and the exchange error terms in case of imperfect control are included in \cite{note}. We also note that because $0\le\pi+\phi\le2\pi$ on the rhs of Eq.~\eqref{eq:hfcorrdU12} to ensure positive pulse durations, the angle in its lhs must be restricted to $-\pi\le\phi\le\pi$.

\begin{figure}
  \includegraphics[width=0.9\columnwidth]{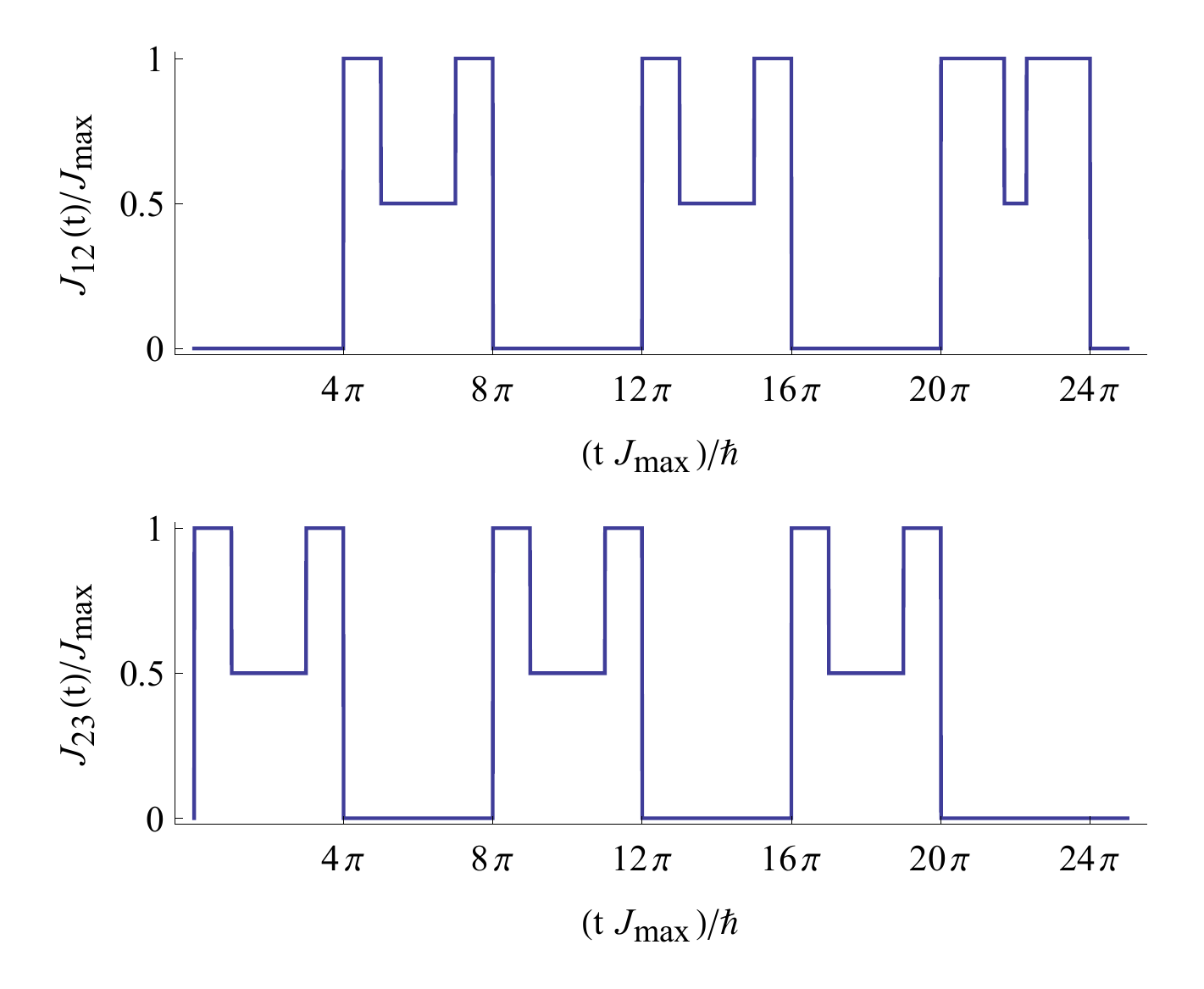}
\caption{Example pulse sequence for corrected rotation achieving $R_{12}(0.7\pi)$.}
\label{fig:E12Pulses}
\end{figure}

\begin{figure}
  \includegraphics[width=0.9\columnwidth]{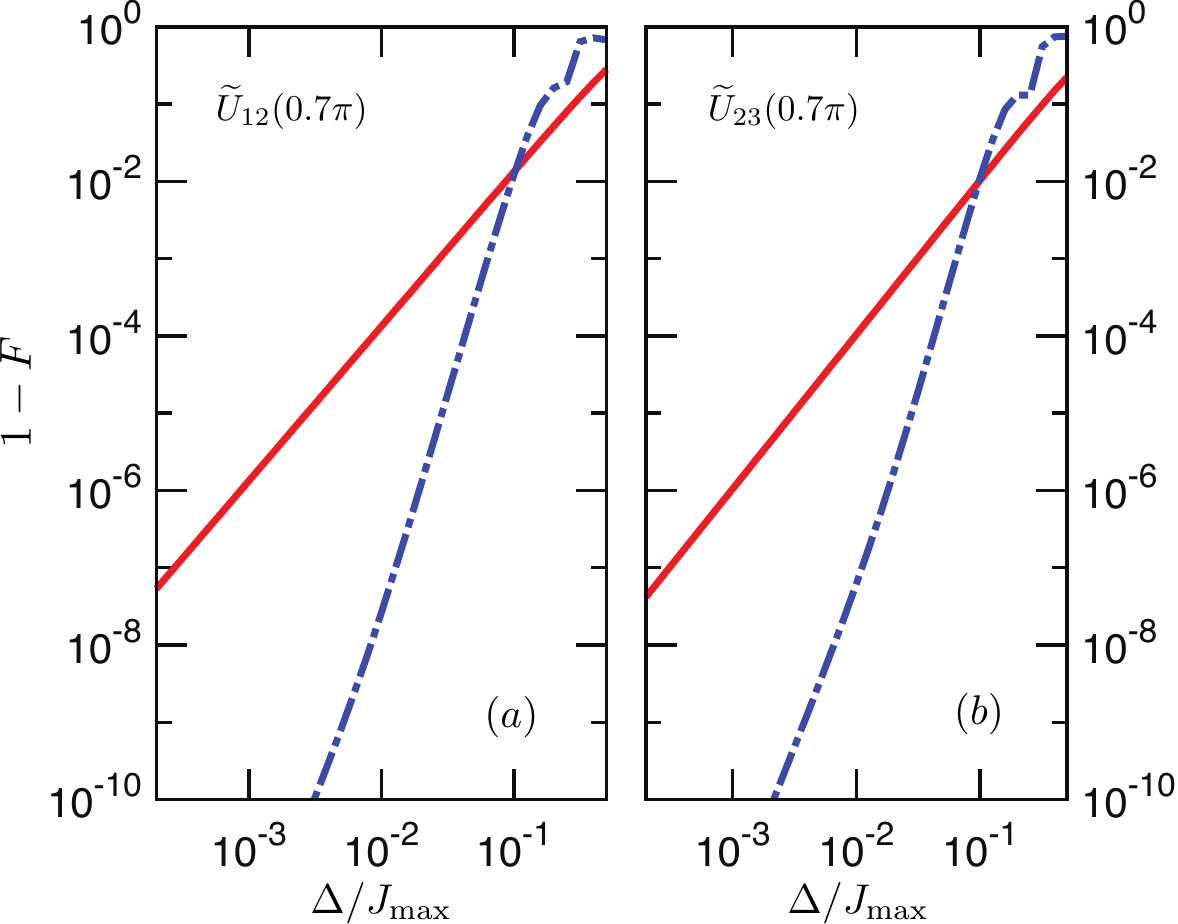}
  \caption{Gate infidelity of na\"{i}ve (solid) and hyperfine-corrected (dot-dashed lines) rotations, assuming perfect control. The rotations implemented are, (a) $\widetilde{U}_{12}(0.7\pi)$, and (b) $\widetilde{U}_{23}(0.7\pi)$. Here we have taken $\Delta_{12}=\Delta_{\overline{12}}=\Delta$. These assumptions are for display purposes only and will not affect the generality of our result.}\label{fig:dherror}
\end{figure}

In Fig.~\ref{fig:E12Pulses} we show an example of pulse sequences achieving $R_{12}(0.7\pi)$. For $R_{23}(0.7\pi)$ one simply needs to swap the upper and lower panels. (Although we have used square pulses, finite rise times can be accommodated \cite{note}.)
Fig.~\ref{fig:dherror} shows the gate infidelity of these rotations for a range of hyperfine fields. The na\"{i}ve pulse is taken to be the fastest possible with the same constraint on $J$. As one would expect, the infidelity in the uncorrected sequence is roughly second-order while that of the corrected sequence appears to be fourth-order in $\Delta_{12}$ and $\Delta_{\overline{12}}$. Our pulse sequence delivers a substantial error reduction, allowing it to tolerate a much wider range of hyperfine error within the quantum error correction threshold.

We now proceed to the cancellation of the exchange noise. This is not trivial because the strategy of Eq.~\eqref{eq:3piecePhi} does not apply due to the peculiar control-dependent form of the exchange noise. In fact, any similar form involving a cascade of $J_{12}$ rotations, $\prod_i U_{12}(J_i,\phi_i)$, could only possibly work if the pulse form is chosen based on exact \textit{a priori} knowledge of $g(J)$, or in other words, the exact functional dependence of $J\left(\epsilon\right)$.  Since that is a sample-dependent property, it is far better to consider a form that will work more generally. Therefore, in order to find a $J_{12}$ pulse sequence which cancels exchange noise for any $g(J)$, we must invoke both $J_{12}$ and $J_{23}$ rotations -- complicating matters by introducing another independent error source, $\delta \epsilon_{23}$ -- and assign rotation angles to them such that all error terms are canceled simultaneously. To simplify the task (no doubt at the cost of pulse length), we can build all rotations from the hyperfine-corrected building blocks $\widetilde{U}_{12/23}(\phi)$ so that we can focus on the exchange noise only. The algebra is rather involved, but in this way we have found a sequence which leads to a cancellation of exchange noise for both axes. We first define an identity up to exchange error terms
\begin{multline}
\widetilde{I}_{12}(\phi_1,\phi_2)=\widetilde{U}_{12}(-\phi_2)\widetilde{U}_{23}(-\phi_1)
\\
\times\left[\widetilde{U}_{12}(\pi)\right]^2 \widetilde{U}_{23}(\phi_1) \widetilde{U}_{12}(\phi_2),\label{eq:djidentity}
\end{multline}
and then we can express the sequence that corrects both exchange and hyperfine errors to leading order as
\begin{multline}
\widetilde{\widetilde{U}}_{12}(\phi)
\equiv \widetilde{U}_{12}(\phi)\widetilde{I}_{12}(\phi_1,\phi_2) \widetilde{I}_{12}(-\phi_1,-\phi_2)
\\
= R_{12}(\phi) + \mathcal{O}\left[\left(\Delta_{12}+\Delta_{\overline{12}}+\delta\epsilon_{12}+\delta\epsilon_{23}\right)^2\right],\label{eq:djseq}
\end{multline}
where the parameters are given by \cite{note}
\begin{align}
\phi_1&=\arccos\left(-\frac{\pi +\phi }{3\pi }\right),\label{eq:phi1}\\
\phi_2&=\arctan\left(\frac{4 \pi +\phi }{2 \sqrt{8 \pi ^2-2 \pi  \phi -\phi ^2}}\right).\label{eq:phi2}
\end{align}
We note that for any $\phi\in[-\pi,\pi]$, Eqs.~\eqref{eq:phi1} and \eqref{eq:phi2} are well defined and the resulting $\phi_1$ and $\phi_2$ are both within the prescribed range $[-\pi,\pi]$. Therefore Eqs.~\eqref{eq:3piecePhi}--\eqref{eq:phi2} complete the core results of this paper.

\begin{figure}
  \includegraphics[width=0.9\columnwidth]{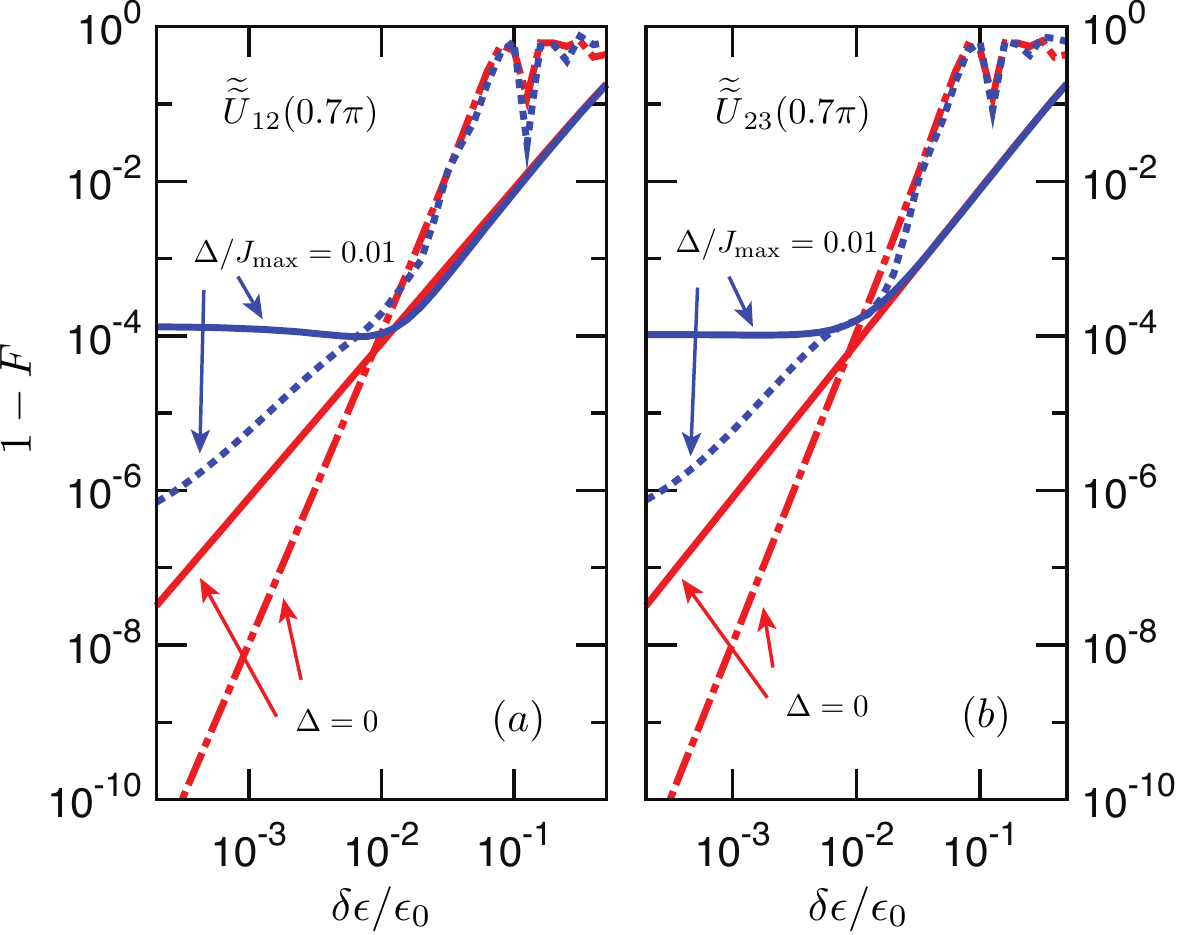}
  \caption{Gate infidelity of na\"{i}ve (solid) rotations and  rotations with both exchange and hyperfine noise corrected (dot-dashed and dotted lines). The rotations implemented are, (a) $\widetilde{\widetilde{U}}_{12}(0.7\pi)$, and (b) $\widetilde{\widetilde{U}}_{23}(0.7\pi)$. Here we have taken $g_{12}(J)=g_{23}(J)=J/\epsilon_0$. We have also taken $\Delta_{12}=\Delta_{\overline{12}}=\Delta$ and $\delta\epsilon_{12}=\delta\epsilon_{23}=\delta\epsilon$. Again, these assumptions are for display purposes only.}\label{fig:Jerror}
\end{figure}

In Fig.~\ref{fig:Jerror} we show the infidelity of gates performing $R_{12}(0.7\pi)$ and $R_{23}(0.7\pi)$. The red lines are for the case without the hyperfine error; while blue lines are for hyperfine error $\Delta_{12}=\Delta_{\overline{12}}=0.01J_\mathrm{max}$. The corrected pulse clearly delivers improvement to the na\"ive pulses for small charge noise. The improvement is appreciable only at very small $\delta \epsilon$ because, although the first order error is completely cancelled, one is left with a large coefficient in the next order error. This coefficient could likely be reduced by further optimization of the pulse. Nevertheless, the relevant parameter regime for typical experiments is estimated to be that of small charge noise dominated by hyperfine noise \cite{Mehl12, Medford13}, which is the regime where the fully-corrected pulse could already be useful in its current form.

The cost for the error cancellation is that we have made the pulse sequence long. For the hyperfine-corrected sequence, typically $18\pi$ of rotation is needed. For the sequences robust against both hyperfine error and the exchange error, one would need $\sim230\pi$ of rotations. Although this number seems challenging, recent experiment \cite{Medford13} has demonstrated $\sim160\pi$ coherent rotations within $\sim2$ ns, which is well below even their $T_2^*$ value (a lower bound on $T_2$) of $\sim100$ns. This suggests that our hyperfine-corrected pulse sequence can be immediately implemented in experiments, while even the much longer, fully-corrected sequence could be within practicality. Moreover, for realistic situations one usually controls $J_{12}$ and $J_{23}$ with one single detuning, which can be used to further optimize the sequence. We emphasize that the pulse sequence discussed in this paper is completely general, canceling $\delta\epsilon_{12}$ and $\delta\epsilon_{23}$ independently and making no assumption whatsoever on how the exchange depends on detuning. Further optimization would require input of this information and could vary from system to system, requiring a case-by-case analysis.

It is interesting to consider error cancellation beyond the first order. In \cite{note} we demonstrate that in the absence of exchange error, hyperfine error can be corrected up to second order by cascading the pulse of Eq.~\ref{eq:3piecePhi}. However, pushing to higher orders does not necessarily improve the performance of the pulses, as shown in \cite{Wang12}. We therefore speculate that it would be more useful optimizing the pulse sequences presented here, rather than canceling higher orders of errors by a much longer sequence.

In conclusion, we have presented analytical composite pulse sequences for arbitrary rotations of an exchange-only qubit, which are simultaneously robust against both quasistatic nuclear noise and charge noise.  The form of our sequences satisfies the physical constraints of the exchange-only system, does not assume \textit{a priori} knowledge of the exact functional dependence of the exchange interaction on detuning, and is realistic for current experimental implementation. We therefore believe that this control method will be immensely useful to on-going experimental efforts towards scalable arrays of qubits with all-electrostatic control.

Research was supported by LPS and by the Office of the Director of National Intelligence, Intelligence Advanced Research Projects Activity (IARPA), through the Army Research Office grant W911NF-12-1-0354.


\onecolumngrid
\newpage

\vspace{1cm}
\begin{center}
{\bf\large Supplementary material}
\end{center}
\vspace{0.5cm}

\setcounter{secnumdepth}{3}  
\setcounter{equation}{0}
\setcounter{figure}{0}
\renewcommand{\theequation}{S-\arabic{equation}}
\renewcommand{\thefigure}{S\arabic{figure}}
\renewcommand\figurename{Supplementary Figure}
\newcommand\Scite[1]{[S\citealp{#1}]}

\makeatletter \renewcommand\@biblabel[1]{[S#1]} \makeatother


\section{Gell-Mann matrices}

Here for convenience we give the explicit forms of the Gell-Mann matrices \Scite{Georgi99} used in the paper:
\begin{equation}
\begin{gathered}
\lambda_1=\left(
\begin{array}{ccc}
 0 & 1 & 0 \\
 1 & 0 & 0 \\
 0 & 0 & 0
\end{array}
\right),\ \lambda_2=\left(
\begin{array}{ccc}
 0 & -i & 0 \\
 i & 0 & 0 \\
 0 & 0 & 0
\end{array}
\right),\ \lambda_3=\left(
\begin{array}{ccc}
 1 & 0 & 0 \\
 0 & -1 & 0 \\
 0 & 0 & 0
\end{array}
\right),\ \lambda_4=\left(
\begin{array}{ccc}
 0 & 0 & 1 \\
 0 & 0 & 0 \\
 1 & 0 & 0
\end{array}
\right),\\
\lambda_5=\left(
\begin{array}{ccc}
 0 & 0 & -i \\
 0 & 0 & 0 \\
 i & 0 & 0
\end{array}
\right),\
\lambda_6=\left(
\begin{array}{ccc}
 0 & 0 & 0 \\
 0 & 0 & 1 \\
 0 & 1 & 0
\end{array}
\right),\ \lambda_7=\left(
\begin{array}{ccc}
 0 & 0 & 0 \\
 0 & 0 & -i \\
 0 & i & 0
\end{array}
\right),\ \lambda_8=\frac{1}{\sqrt{3}}\left(
\begin{array}{ccc}
 1 & 0 & 0 \\
 0 & 1 & 0 \\
 0 & 0 & -2
\end{array}
\right).
\end{gathered}
\end{equation}

\section{Analytical forms of pulse sequences}

In this section we present the detailed analytical form of the pulse sequences presented in the main text of the paper.

A na\"ive rotation with $J_{12}$, $U_{12}(J,\phi)$, defined in Eq.~\eqref{eq:U12def}, has first order error as
\begin{equation}\begin{split}
U_{12}(J,\phi)=R_{12}(\phi)\Bigg\{&I-i\frac{\sin\phi}{2 \sqrt{3}J} \Delta_{12}\lambda_1
+i\frac{\cos\phi-1}{2\sqrt{3}J}\Delta_{12}\lambda_2
+\left(-\frac{i \phi }{3J}\Delta_{\overline{12}}+\frac{i\phi g_{12}\left(J\right)}{2J}\delta\epsilon_{12}\right)\lambda_3
-i\frac{\sin\phi}{\sqrt{6}J}\Delta_{12}\lambda_4\\
&+i\frac{\cos\phi-1}{\sqrt{6}J}\Delta_{12}\lambda_5
-i\frac{\sqrt{2}}{3}\frac{\phi}{J}\Delta_{\overline{12}}\lambda_6
+i\frac{\phi g_{12}\left(J\right)}{2\sqrt{3}J}\delta\epsilon_{12}\lambda_8+{\cal O}\left[\left(\Delta_{12}+\Delta_{\overline{12}}+\delta\epsilon_{12}+\delta\epsilon_{23}\right)^2\right]\Bigg\}.
\end{split}\end{equation}

The expression for $U_{23}(J,\phi)$ is similar but lengthy, therefore we will not present it here.

The three-piece pulse sequences [Eq.~\eqref{eq:3piecePhi}] forming the building blocks of the hyperfine-corrected rotation are
\begin{equation}\begin{split}
U_{12}'(J,\phi)=e^{-\frac{2i\pi}{3}}R_{12}(\phi)\Bigg\{&I+
\left[-\frac{4i\pi}{3J}\Delta_{\overline{12}}+\frac{i}{J}\left[(2 \pi -\phi ) g_{12}\left(\frac{J}{2}\right)
+\phi g_{12}\left(J\right)\right]\delta\epsilon_{12}\right]\lambda_3
-i\frac{4\sqrt{2}\pi}{3J}\Delta_{\overline{12}}\lambda_6\\
&+\frac{i}{\sqrt{3}J}\left[(2 \pi -\phi ) g_{12}\left(\frac{J}{2}\right)+\phi g_{12}\left(J\right)\right]\delta\epsilon_{12}\lambda_8
+{\cal O}\left[\left(\Delta_{12}+\Delta_{\overline{12}}+\delta\epsilon_{12}+\delta\epsilon_{23}\right)^2\right]\Bigg\},
\end{split}\label{suppleq:U123piece}\end{equation}
and
\begin{equation}\begin{split}
U_{23}'(J,\phi)=
e^{-\frac{2 i \pi }{3}}R_{23}(\phi)
\Bigg\{&I+\left[-\frac{\sqrt{3}\pi i}{2J}\Delta_{12}+\frac{i \pi }{\sqrt{3}J}\Delta_{\overline{12}}
+\frac{\sqrt{3}i}{2J}\left[(2 \pi -\phi ) g_{23}\left(\frac{J}{2}\right)+\phi  g_{23}\left(J\right)\right]\delta\epsilon_{23}\right]\lambda_1\\
&+\left[\frac{i \pi }{2J}\Delta_{12}-\frac{i \pi }{3J}\Delta_{\overline{12}}-\frac{i}{2J}\left[(2 \pi -\phi ) g_{23}\left(\frac{J}{2}\right)+\phi  g_{23}\left(J\right)\right]\delta\epsilon_{23}\right]\lambda_3\\
&+\frac{i\pi}{J}\left(-\sqrt{\frac{3}{2}}\Delta_{12}+\sqrt{\frac{2}{3}}\Delta_{\overline{12}}\right)\lambda_4
+\frac{i\pi}{J}\left(\frac{\Delta_{12}}{\sqrt{2}}-\frac{\sqrt{2}}{3}\Delta_{\overline{12}}\right)\lambda_6\\
&+\frac{i}{\sqrt{3}J}\left[(2 \pi -\phi ) g_{23}\left(\frac{J}{2}\right)+\phi  g_{23}\left(J\right)\right]\delta\epsilon_{23}\lambda_8+{\cal O}\left[\left(\Delta_{12}+\Delta_{\overline{12}}+\delta\epsilon_{12}+\delta\epsilon_{23}\right)^2\right]\Bigg\}.
\end{split}\label{suppleq:U233piece}\end{equation}
Here, the overall phase factor $e^{-\frac{2 i \pi }{3}}$ indicates that the actual rotation angle is $2\pi+\phi$. Note that the error terms related to $\Delta_{12}$ and $\Delta_{\overline{12}}$ do not have any $\phi$ dependence.  Moreover, the $\lambda_2$ term vanishes in both Eqs.~\eqref{suppleq:U123piece} and \eqref{suppleq:U233piece}. Therefore when we combine them as in Eq.~\eqref{eq:hfcorrdU12}, all hyperfine-induced errors cancel.

The hyperfine-corrected rotations, defined in Eq.~\eqref{eq:hfcorrdU12}, are
\begin{equation}\begin{split}
\widetilde{U}_{12}(\phi)=R_{12}(\phi)\Bigg\{&I
-i\phi\left[g_{12}\left(\frac{1}{2}\right)-g_{12}\left(1\right)\right]\delta\epsilon_{12}\lambda_3\\
&+\left[\frac{i}{\sqrt{3}}\left[(3 \pi -\phi ) g_{12}\left(\frac{1}{2}\right)+(3 \pi +\phi ) g_{12}\left(1\right)\right]\delta\epsilon_{12}+i \sqrt{3} \pi  \left[g_{23}\left(\frac{1}{2}\right)+g_{23}\left(1\right)\right]\delta\epsilon_{23}\right]\lambda_8\\
&+{\cal O}\left[\left(\Delta_{12}+\Delta_{\overline{12}}+\delta\epsilon_{12}+\delta\epsilon_{23}\right)^2\right]\Bigg\},
\end{split}\label{suppleq:U12hfcorrd}\end{equation}
and
\begin{equation}\begin{split}
\widetilde{U}_{23}(\phi)=R_{23}(\phi)\Bigg\{&I
-\frac{\sqrt{3}i\phi}{2}\left[g_{23}\left(\frac{1}{2}\right)-g_{23}\left(1\right)\right]\delta\epsilon_{23}\lambda_1+\frac{i\phi}{2}\left[g_{23}\left(\frac{1}{2}\right)-g_{23}\left(1\right)\right]\delta\epsilon_{23}\lambda_3\\
&+\left[i\sqrt{3}\pi\left[g_{12}\left(\frac{1}{2}\right)+g_{12}\left(1\right)\right]\delta\epsilon_{12}+\frac{i}{\sqrt{3}}\left[(3 \pi -\phi ) g_{23}\left(\frac{1}{2}\right)+(3 \pi +\phi ) g_{23}\left(1\right)\right]\delta\epsilon_{23}\right]\lambda_8\\
&+{\cal O}\left[\left(\Delta_{12}+\Delta_{\overline{12}}+\delta\epsilon_{12}+\delta\epsilon_{23}\right)^2\right]\Bigg\}.
\end{split}\label{suppleq:U23hfcorrd}\end{equation}
(Here we took $J=1$.) We see that all first-order dependences on $\Delta_{12}$ and $\Delta_{\overline{12}}$ vanish. In case of perfect control, this sequence would be error-free in the first order.

The exchange-noise-corrected pulse sequence, building from Eqs.~\eqref{suppleq:U12hfcorrd} and \eqref{suppleq:U23hfcorrd} according to Eqs.~\eqref{eq:djidentity} and \eqref{eq:djseq}, are
\begin{equation}\begin{split}
\widetilde{\widetilde{U}}_{12}(\phi)=e^{\frac{2 i \pi }{3}}R_{12}(\phi)\Bigg\{&I
-i \sqrt{3}\pi\left[g_{12}\left(\frac{1}{2}\right)-g_{12}\left(1\right)\right]\left[(\cos\phi_1-1)\cos\phi_2+2\sin\phi_1 \sin\phi_2\right]\delta\epsilon_{12} \lambda_1\\
&-i (\pi +\phi +3 \pi  \cos\phi_1) \left[g_{12}\left(\frac{1}{2}\right)-g_{12}\left(1\right)\right]\delta\epsilon_{12} \lambda_3\\
&+\left[\frac{i}{\sqrt{3}}\left[(35 \pi -\phi ) g_{12}\left(\frac{1}{2}\right)+(43 \pi +\phi ) g_{12}\left(1\right)\right]\delta\epsilon_{12}+13\sqrt{3}i\pi\left[g_{23}\left(\frac{1}{2}\right)+g_{23}\left(1\right)\right]\delta\epsilon_{23}\right]\lambda_8\\
&+{\cal O}\left[\left(\Delta_{12}+\Delta_{\overline{12}}+\delta\epsilon_{12}+\delta\epsilon_{23}\right)^2\right]\Bigg\},
\end{split}\label{suppleq:U12djcorrd}\end{equation}
and
\begin{equation}\begin{split}
\widetilde{\widetilde{U}}_{23}(\phi)=e^{\frac{2 i \pi }{3}} R_{23}&(\phi)\Bigg\{I\\
&-\frac{\sqrt{3}i}{2}\left[g_{23}\left(\frac{1}{2}\right)-g_{23}\left(1\right)\right][\pi +\phi -\pi  \cos\phi_2+\pi  \cos\phi_1 (3+\cos\phi_2)+2 \pi  \sin\phi_1 \sin\phi_2]\delta\epsilon_{12} \lambda_1\\
&+\frac{i}{2}\left[g_{23}\left(\frac{1}{2}\right)-g_{23}\left(1\right)\right][\pi +\phi +3 \pi  (\cos\phi_1+\cos\phi_2-\cos\phi_1 \cos\phi_2-2 \sin\phi_1 \sin\phi_2)]\delta\epsilon_{12} \lambda_3\\
&+\left[13\sqrt{3}i\pi\left[g_{12}\left(\frac{1}{2}\right)+g_{12}\left(1\right)\right]\delta\epsilon_{12}+\frac{i}{\sqrt{3}}\left[(35 \pi -\phi ) g_{23}\left(\frac{1}{2}\right)+(43 \pi +\phi ) g_{23}\left(1\right)\right]\delta\epsilon_{23}\right]\lambda_8\\
&+{\cal O}\left[\left(\Delta_{12}+\Delta_{\overline{12}}+\delta\epsilon_{12}+\delta\epsilon_{23}\right)^2\right]\Bigg\}.
\end{split}\label{suppleq:U23djcorrd}\end{equation}
We first examine Eq.~\eqref{suppleq:U12djcorrd}. To make the undesired $\delta\epsilon_{12}$ and $\delta\epsilon_{23}$ terms vanish, we need
\begin{align}
\pi +\phi +3 \pi  \cos\phi_1&=0\\
(\cos\phi_1-1)\cos\phi_2+2\sin\phi_1 \sin\phi_2&=0.
\end{align}
One can readily verify that the parameters given in the main text, Eqs.~\eqref{eq:phi1} and \eqref{eq:phi2}, satisfy these requirements. It is also straightforward to verify that the same $\phi_1$ and $\phi_2$ will cancel the undesired $\delta\epsilon_{12}$ and $\delta\epsilon_{23}$ terms in Eq.~\eqref{suppleq:U23djcorrd}. We remark here that this error cancellation scheme does not assume any particular form of $g(J)$.
Interestingly, one may note that after all first order error terms are cancelled, the remaining $\lambda_8$ terms in Eqs.~\eqref{suppleq:U12djcorrd} and \eqref{suppleq:U23djcorrd} are identical  when indices ${12}$ and ${23}$ are interchanged. This reflects the fact that the $|Q\rangle$ state is completely symmetric with respect to the three spins.

\section{Accounting for finite rise time}

\begin{figure}[]
  \includegraphics[width=1.1\columnwidth]{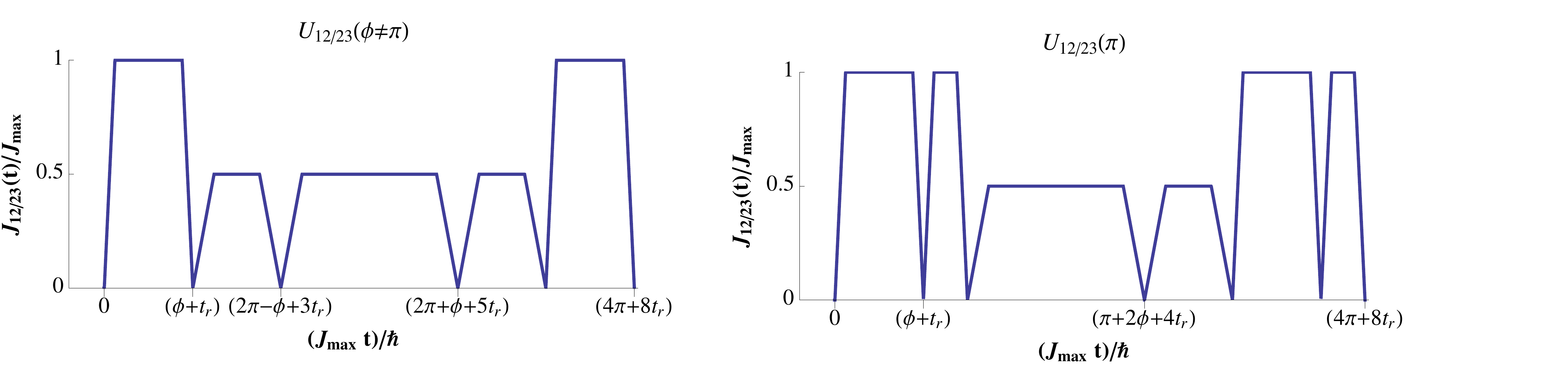}
\caption{Example sequences for $U'_{12/23}(J,\phi\neq\pi,t_r)$ and $U'_{12/23}(J,\pi,t_r)$, to replace Eq.~\eqref{eq:3piecePhi} in situations where finite rise times cannot be neglected. Here $\phi=0.7\pi$ and $t_r=0.3$, where $t_r$ is the rise time scaled by $J_{max}/\hbar$.}
\label{fig:U12FiniteRise}
\end{figure}

So far our analysis has assumed that exchange strengths can be varied instantaneously. In a laboratory implementation though there must be a finite rise or fall time associated with each change in $J$. Here we show that even when these effects are taken into account, our first main result still holds. That is, with only slight modifications Eq.~\eqref{eq:hfcorrdU12} still cancels all hyperfine-induced errors to first order.

As noted in the main text, Eq.~\eqref{eq:phitot2} applies regardless of pulse shape. Thus to construct an error-cancelling sequence in the presence of finite rise times we need only find a new implementation for the $U'_{12/23}(J,\phi)$'s that still satisfies $a_2=b_2$ in that regime. To that end we revisit the intuition that Eq.~\eqref{eq:3piecePhi} can result in $a_2=b_2=0$ by spending equal time intervals at each pair of diametrically opposite points on its path around the Bloch sphere. Now, finite rise times force the qubit to spend extra time at the beginning and ending portions of pulses, due to the steady increase or decrease in the value of $J$. However, the error induced at such a point can be exactly cancelled by forcing the qubit to spend the same amount of extra time at the corresponding point directly opposite from it across the rotation axis, i.e. $\pi$ radians of rotation away. This can be accomplished by forcing $J$ to zero at judiciously-chosen points, with the correct fall and rise times, during the $U'_{12/23}(J,\phi)$ implementation.

Let $U_{12/23}(J,\phi,t_r)$ denote an operation in which $J\left(t\right)$ is pulsed from zero to $J$, held at $J$, then returned to zero. The rise and fall times are both equal to $t_r$, and the duration is chosen such that, including the ramping portions, the total angle of rotation accomplished is $\phi$. Our result here requires a separate implementation for $U'_{12/23}(J,\pi,t_r)$ and $U'_{12/23}(J,\phi\neq\pi,t_r)$, each of which can be written as:

\begin{equation}
U'_{12/23}(J,\phi,t_r)=U_{12/23}(J,\phi,t_r)U_{12/23}\left(\frac{J}{2},\pi-\phi,2t_r\right)U_{12/23}\left(\frac{J}{2},\phi,2t_r\right) U_{12/23}\left(\frac{J}{2},\pi-\phi,2t_r\right)U_{12/23}(J,\phi,t_r)
\end{equation}
and
\begin{multline}
U'_{12/23}(J,\pi,t_r)=U_{12/23}(J,\phi,t_r)U_{12/23}\left(J,\pi-\phi,t_r\right)U_{12/23}\left(\frac{J}{2},\phi,2t_r\right) \\ \times U_{12/23}\left(\frac{J}{2},\pi-\phi,2t_r\right)U_{12/23}(J,\phi,t_r)U_{12/23}\left(J,\pi-\phi,t_r\right)
\end{multline}

The rise time corrected implementations are plotted in Supplementary Figure \ref{fig:U12FiniteRise}. They satisfy $a_2=b_2=0$, and hence result in first-order correction when used as replacements for Eq.~\eqref{eq:3piecePhi}. In the limit of $t_r$ approaching zero these implementations are identical to those of Eq.~\eqref{eq:3piecePhi}. Note that when applying these pulses we must constrain $-\pi\leq\phi\leq 0$ in Eq.~\eqref{eq:hfcorrdU12}. A similar result can be derived for rotations with the range $0\leq\phi\leq\pi$, or the sequence given here may be applied successively to reach the desired angle.

\section{second-order error cancellation for hyperfine noise}

In this section we demonstrate that if we have perfect control, then the hyperfine noise can be cancelled up to second order. Recall the construction of Eq.~\eqref{eq:hfcorrdU12}. If we find a composite pulse, $\breve{U}_{12}'(J, \phi)$, the error of which is independent of $\phi$ up to the second order of $\Delta_{12}$ and $\Delta_{\overline{12}}$, then the sequence
\begin{align}
\breve{U}_{12}(\phi)\equiv \breve{U}_{12}'(J,\pi+\phi)U_{23}'(J,\pi)\breve{U}_{12}'(J,\pi)U_{23}'(J,\pi)\breve{U}_{12}'(J,\pi)U_{23}'(J,\pi)\label{suppleq:2ndordersequence}
\end{align}
would perform a rotation robust against hyperfine noise up to second order. Note that we only need a ``first-order'' corrected $U_{23}'(J,\pi)$ in Eq.~\eqref{suppleq:2ndordersequence} [\emph{not} $\breve{U}_{23}'(J,\pi)$] since its error is of course $\phi$ independent (only $\pi$ rotation is needed) and its error on $\lambda_2$ can be verified to remain zero up to the second order. Also, the strength of the control field in $U_{23}'(J,\pi)$ and $\breve{U}_{12}'(J, \phi)$ does not have to be equal, as already noted in the main text.

We have found $\breve{U}_{12}'(J, \phi)$ having the desired property by cascading $U_{12}'(J, \phi)$ defined in Eq.~\eqref{eq:3piecePhi} as
\begin{align}
\breve{U}_{12}'(J,\phi)\equiv U_{12}'\left(J,\frac{\phi}{3}\right)U_{12}'\left(J,\frac{\phi}{3}\right)U_{12}'\left(\frac{J}{2},2\pi-\frac{\phi}{3}\right)U_{12}'\left(J,\frac{\phi}{3}\right)U_{12}'\left(J,\frac{\phi}{3}\right).\label{suppleq:Ubreveprimestack}
\end{align}
Direct algebra shows
\begin{equation}
\begin{split}
\breve{U}_{12}'(J,\phi)&=R_{12}(\phi)
\Bigg[\left(1-\frac{64\pi ^2}{J^2}\Delta_{\overline{12}}^2\right)I+\left(-\frac{8 i \pi }{J} \Delta_{\overline{12}}+\frac{8 i \pi }{J^2} \Delta_{12}^2+\frac{32 \pi ^2}{J^2}\Delta_{\overline{12}}^2\right)\lambda_3\\
&\qquad+\left(-\frac{8 i \sqrt{2} \pi  }{J}\Delta_{\overline{12}}-\frac{4 i \sqrt{2} \pi }{J^2} \Delta_{12}^2+\frac{32 \sqrt{2} \pi ^2}{J^2}\Delta_{\overline{12}}^2\right)\lambda_6+\left(\frac{4 i \sqrt{3} \pi  }{J^2}\Delta_{12}^2\right)\lambda_8\Bigg],
\end{split}\label{suppleq:Ubreveprimedef}
\end{equation}
and it is obvious that there is no $\phi$ dependence in the error terms on the rhs of Eq.~\eqref{suppleq:Ubreveprimedef}. One can then calculate Eq.~\eqref{suppleq:2ndordersequence} as
\begin{equation}
\breve{U}_{12}(\phi)=R_{12}(\phi)\Bigg\{I+\frac{3\sqrt{3}i\pi}{2}\left(\frac{33}{4}\Delta_{12}^2+\Delta_{12}\Delta_{\overline{12}}+\Delta_{\overline{12}}^2\right)\lambda_8+{\cal O}\left[\left(\Delta_{12}+\Delta_{\overline{12}}\right)^3\right]\Bigg\},
\label{suppleq:2ndordercorrectedseq}
\end{equation}
where the hyperfine noise induced error is cancelled up to the second order (we have taken $J=1$). For $\breve{U}_{23}(\phi)$ corrected to the same order, one simply interchanges the $12$ and $23$ indices in Eqs.~\eqref{suppleq:2ndordersequence} and \eqref{suppleq:Ubreveprimestack}. We also note that for all rotation angles to be guaranteed non-negative, we require $(\pi+\phi)/3\ge0$ and $2\pi-(\pi+\phi)/3\ge0$ [from the rhs of Eq.~\eqref{suppleq:Ubreveprimestack}], which is obviously satisfied for any $\phi\in[-\pi,\pi]$.

\section{Definition of fidelity}

In Figs.~\ref{fig:dherror} and \ref{fig:Jerror} we characterize the error in terms of the infidelity (one minus the fidelity, $F$). The fidelity has been discussed in Refs.~\Scite{Bowdrey02} and \Scite{Cabrera07}. However, due to presence of the leaked state $|Q\rangle$, our evolution operator is not trace-preserving. Therefore Eq.~(7) of Ref. \Scite{Bowdrey02} must be extended as
\begin{equation}
 F= \frac{1}{4}\mathrm{Tr}\left(V\widetilde{\lambda}_0V^\dagger U_f\widetilde{\lambda}_0U_f^\dagger+\frac{1}{3}\sum_{j=1}^3V\lambda_jV^\dagger U_f\lambda_jU_f^\dagger\right) ,
\end{equation}
where $V$ is the desired (noiseless) operation, $U_f$ is the actual time evolution at the conclusion of composite pulse sequence, and
\begin{equation}
\widetilde{\lambda}_0=\left(
\begin{array}{ccc}
 1 & 0 & 0 \\
 0 & 1 & 0 \\
 0 & 0 & 0
\end{array}
\right).
\end{equation}

\end{document}